\documentstyle[twocolumn,aps,prl,psfig]{revtex} 
 
\def\wsep{$W_{sep}$} 
\def\al2o3{Al$_2$O$_3$}
\def\AO{$A_mO_n$}
 
\begin{document} 
\draft 
\title{ Equilibrium and adhesion of Nb/sapphire: the effect
of oxygen partial pressure} 
\author{Iskander G. Batirev$^{1,2}$, Ali Alavi$^1$ and Michael W. Finnis$^1$ 
 } 
\address{$^1$Atomistic Simulation Group, School of Mathematics and Physics, 
The Queen's University of Belfast, Belfast BT7 1NN, Northern Ireland, UK} 
\address{$^2$The University of Texas at Austin, Department of Physics,Austin, 
Texas 78712-1081} 
 
\maketitle

\begin{abstract}

We derive a formula, useful for first-principles calculations, which
relates the free energy of an oxide/metal interface to the free
energies of  surfaces and the work of separation of the interface.  We
distinguish the latter {\it mechanical} quantity from the {\it
thermodynamic} work of adhesion, and we describe explicitly how both
may be calculated.  Our formulae for interfacial and surface energies
are cast in terms of quantities which can be calculated or looked up in
tables, and include as additional parameters the ambient temperature
and partial pressure of oxygen $P_{O_2}$. From total energy
calculations for the Nb(111)/$\alpha-$\al2o3 (0001) interface, free Nb
and \al2o3 surfaces, we obtain firstly numerical estimates of the works
of separation, which are independent of  $P_{O_2}$. We then obtain
surface energies, interfacial energies and the equilibrium work of
adhesion as a function of $P_{O_2}$.

\end{abstract} 
 
\date{Today} 
\section{Introduction}

Oxide-metal interfaces continue to be studied intensively because of
the many ways in which they are of commercial and scientific
importance.  Applications range from the nanoscale in microelectronics
packaging to the macroscale engineering of thermal barrier coatings or
the formation of protective scales. The science of these interfaces has
been addressed in volumes of conference papers and reviews
\cite{Ruehle1,Ruehle2,Howe1,Howe2}.  There are also reviews in the
literature \cite{Noguera1,Finnis} which specifically address the
theoretical questions about the nature of the bonding at these
interfaces, such as: what determines the site-preference of metal atoms
on the oxide surface; whether the bonding can be thought of as
predominantly covalent or metallic and how to quantify these concepts;
whether a simple classical image model can be used to interpret the
bonding; what is the strength of adhesion of metal to oxide.  The basis
for answering these questions is to have reliable calculations of the
electronic structure and total energy of particular surfaces and
interfaces. Such calculations came of age over the past ten years or so
with the use of first-principles methods. These  mainly apply density
functional theory (DFT) and the local density approximation (LDA)
\cite{HK,KS}, which are the basis of the calculations we shall report
here. Hartree-Fock calculations are also feasible and have been applied
effectively to the Ag/MgO interface~\cite{Heifets}, although they tend
to be more expensive than DFT for larger systems. Since the reviews
cited, there have been numerous applications of DFT to study bonding in
the initial stages of deposition of metal on oxide, with cluster or
multilayer geometries, notably on
MgO~\cite{Goniakowski1,Goniakowski2,Matveev,Musolino,Neyman1,Neyman2,Pacchioni,Tanaka,Yudanov1,Yudanov2,Benedek1,Benedek2},
but to a lesser extent on more complex oxides such as TiO$_2$ \cite{Paxton},
MgAl$_2$O$_4$ \cite{Koestlmeier} and $\alpha-$\al2o3 \cite{Zhao,Kruse2,Finnis1,Verdozzi}.

The bonding of Nb to $\alpha-$\al2o3 has long been a subject for
experimental work, because besides its relevance to electronic
components it offers practical advantages for sample preparation: the
two materials bond strongly \cite{Korn} (anomalously strongly according
to a recent study~\cite{Song}), do not react chemically and have
similar coefficients of thermal expansion. In the orientation
Nb(111)/\al2o3 (0001) there is a lattice mismatch of $<2\%$, allowing
the preparation of a nearly coherent interface (using molecular beam
epitaxy), the atomic structure of which has been studied by high
resolution transmission electron microscopy (HRTEM) and analysed in
detail~\cite{Mayer1,Mayer2,Gutekunst1,Gutekunst2,Gutekunst3}. This
interface was the subject of first-principles calculations which used
periodic boundary conditions, making the reasonable assumption that the
effect of misfit dislocation can be neglected
\cite{Kruse2,Finnis1,Verdozzi}.  Our recent work \cite{Finnis1}
analysed the nature of the bonding in detail by calculating Mulliken
populations and bond orders, concluding that the bonding across the
interface is strongly ionic. The work of separation \wsep\ of the
interface was calculated, and found to be very high: of order
10\,Jm$^{-2}$ when niobium was bonded to the oxygen terminated \al2o3
surface. Lower energy pathways for the cleavage of this interface would
be within the Nb metal or the oxide itself. Two other interfaces were
studied corresponding to the two other possible terminations of bulk
\al2o3 (0001); namely the stoichiometric, aluminium termination (one
layer of aluminium) and the aluminium-rich termination (two layers of
aluminium). HRTEM could not distinguish between the stoichiometric
termination and the oxygen termination, however evidence from electron
energy loss spectroscopy \cite{Bruley} favoured the oxygen
termination.

We point out here that as far as we know the question of which
termination is more stable has not yet been addressed in all the
theoretical work which has been published so far on any oxide-metal
interfaces. The structural predictions have been confined to the
question of the relative displacement of the crystals, parallel and
perpendicular to the interface, and the local relaxations of atoms at
the interface, as well as the energy needed to separate the crystals
\wsep . This has been done for interfaces with different terminations or
local stoichiometry; all calculations were carried out with atoms at
rest ($T=0K$) and minima in the total energy were located as a function
of atomic positions. However, the question as to whether the oxygen
terminated or the aluminium terminated interface is more stable was not
discussed. It is well known that this question can only be answered with
respect to the chemical potentials of the species in the environment
with which the interfaces are in equilibrium, which is normally
characterised by temperature and partial pressure of oxygen
\cite{Shalz}, and the difficulty of relating the quantities accessible
to a first principles calculation to these parameters may have been a reason
for leaving this question to one side. 

The main purpose of our present paper is to show how in fact we are
already able to make predictions of the stability of different
interfaces when they differ not only in structure but also in
composition. With certain simplifying assumptions we show how this can
now be done with little more effort than the calculations which need to
be done to calculate the work of separation, and we present first
results for the Nb/\al2o3 interface. The ingredients of the theory are
the works of adhesion and surface energies.  For these we draw upon the
results reported briefly in \cite{Finnis1}, supplemented by some
further calculations to discuss the case of oxygen on the Nb surface.
The basic theory is outlined in Section II. We derive the equations for
a general $A_{m}O_{n}$ oxide in contact with a metal $B$; it would
be a short step to generalise them still further to an interface
between arbitrary compounds. Essentially the same thermodynamics was
applied by Wang {\it et al}~\cite{scheffler} in calculations of the
surface energy of oxides with different terminations, over a range of
chemical potentials of oxygen; our theory makes the further connection
to the temperature and in particular the {\it pressure} of oxygen,
which are the parameters directly under the control of the
experimentalist. A detailed study of the \al2o3 surface will be
reported elsewhere~\cite{Finnis2}.

The plan of the rest of the paper is as follows. Sections III-V cover
aspects of our total energy calculations which were not dealt with, or
dealt with only briefly, in our Letter \cite{Finnis1}. In Section III
our method of total energy calculation is summarised. In Section IV we
describe the atomic relaxations {\it parallel} to the interfaces which
are generally not commented upon. Although this structural aspect is
not central to the thrust of our paper, it turned out that lateral
relaxations also have a strong part to play in determining the
interplanar relaxations and energies reported previously, and we
therefore describe them for completeness. In Section V we describe and
comment on the results for the work of separation on different planes
and with different terminations of the interface.  Our calculated
interfacial free energies are presented in Section VI and we conclude
in Section VII.

\section{Principles of Calculating Interfacial and Surface Energies} 

Let us consider the interface between metal $B$ and an oxide of metal
$A$ in equilibrium at temperature and pressure $(T,P)$. The
stoichiometric composition of the $A$ oxide is $A_{m}O_{n}$.  We obtain
the definitions of interfacial quantities by referring to the contents
of a periodically repeated supercell of area $S$ parallel to the
interfaces which it may contain. All extensive thermodynamic
quantitites in the following will refer to the contents of such a
supercell. The interfacial energy per unit area, counting the two
interfaces within each supercell, is given by\cite{Cahn}:

\begin{eqnarray} 
\gamma_{int}&=&(G_{int}(T,P)-N_{A}\mu_{A}(T,P)-N_{O}\mu_{O}(T,P)\nonumber\\ 
      &&-N_{B}\mu_{B}(T,P))/2S, \label{gammaint}
\end{eqnarray} 
where $G_{int}$ is the Gibbs energy of the contents of a 
supercell containing two interfaces, $\mu_{A}$, $\mu_{B}$ and $\mu_{O}$  are the chemical potentials of
the three components, and
$N_{A}$, $N_{B}$ and $N_{O}$ are the numbers of atoms of the three components
within  the supercell. The
denominator $2S$ occurs because there are two interfaces in the
supercell, as required by periodic boundary conditions.
Chemical potentials here are per atom rather than per mole, which would be the 
usual convention for macroscopic thermodynamics. Special cases of 
Eqn.~(\ref{gammaint}) are when either the metal $B$ or the oxide is absent
from the supercell, in which cases we recover expressions for the surface energies
of the oxide $\gamma_{AO}$ and the metal $\gamma_B$ respectively:

\begin{equation} 
\gamma_{AO}=(G_{SAO}(T,P)-N_A\mu_{A}(T,P)-N_O\mu_O(T,P))/2S, \label{gammaAO}
\end{equation} 
 \begin{equation} 
\gamma_B=(G_{SB}(T,P)-N_B\mu_B(T,P))/2S. \label{gammaB}
\end{equation} 

The quantities $G_{SAO}$ and $G_{SB}$ are the Gibbs energies of slabs
of oxide and of metal, with free surfaces separated in their respective
supercells by an adequate thickness of vacuum. We have assumed in
Eqn.(\ref{gammaB}) that the metal surface is clean; this will suffice
for a calculation of the energy of the $\it interface$ discussed
below.  However, we can easily consider for example adding a monolayer
of oxygen to the metal surface in the calculation of $G_{SB}(T,P)$.
Since there is no separate oxide phase, the number of oxygen atoms in
the system, $N_O$, now resides on the metal surface. The contribution
$-N_O\mu_O(T,P)$  must be subtracted in the calculation of the
corresponding surface energy just as in the calculation of the
interface energy. We shall in fact make this calculation of an
`oxidised' Nb surface in the course of obtaining the work of separation
of an interface by a pathway which leaves oxygen on the exposed metal
surface.

The motivation for calculating $\gamma_{int}$ is as follows. An
interface between two crystals requires five parameters for its
macroscopic specification, for example three to specify the relative
crystallographic orientation of the materials and two more to specify
the orientation of the interface. We note in passing that a free
surface in contact with vapour or liquid only requires two parameters
to specify its crystallographic orientation.  There is always a large
set of hypothetical interfaces which have the same five macroscopic
parameters  but which differ in their atomic structure and local
stoichiometry. The member of this set which minimises $\gamma_{int}$
for given chemical potentials is the equilibrium interface. So provided
we know the chemical potentials of the components, we could in
principle predict the atomistic structure of the equilibrium interface,
including its local stoichiometry, by evaluating $\gamma_{int}$ for
each member of the set.  In practice, of course, we can only calculate
$\gamma_{int}$ for a small subset of the entire set, and rely on our
experience and intuition, together with experimental information, to
ensure that we have not omitted an important structure. Prior to the
present work we and others have calculated total energies for a number
of structures at 0K, in which the atomic positions are relaxed by
energy minimisation.  The equations to be derived below show
how to go the two important steps further, namely to correct the 0K
information to finite temperature and to take account of local stoichiometry.

The relationship between local stoichiometry and $\gamma_{int}$ is well
known in thermodynamics as the Gibbs adsorption equation, in which
local stoichiometry is measured in terms of {\it excesses} $\Gamma_i$
of one or more components labelled $i$. Our final version of
Eqn.~(\ref{gammaint}) will be in terms of the excess of oxygen at the
interface with respect to the metal $A$, per unit surface area, which is defined as:

\begin{equation} 
\Gamma_O=(N_O-\frac{n}{m}N_A)/2S. \label{excess}
\end{equation}  

This choice of component $i$ is arbitrary; we could equally well work in terms of
$\Gamma_{Al}$, because they are related through

\begin{equation} 
m\Gamma_O + n\Gamma_{Al}=0.
\end{equation} 

We note three further points in connection with excesses. Firstly, each
further component in the system would introduce another excess, each
excess being referred to the same designated component. Secondly,  a
stoichiometric interface is by definition one for which all the
excesses vanish. Finally, since one of the phases is the pure metal
$B$, there can be no excess of the metal $B$.  In particular, if $B$
were in fact also $Al$, thereby reducing the number of components to two, the
interface could not be described as having an excess of $O$ or $Al$. For
more discussion of the thermodynamics of excess quantities the reader is
referred to \cite{Cahn}.

A difficulty up to now has been to calculate the chemical
potentials involved in these equations and, more specifically, to relate
them to given experimental conditions. In the following we show how
Eqn.~(\ref{gammaint}) can be reformulated to relate $\gamma_{int}$
to the partial pressure of oxygen $P_{O_2}$.

First we define the {\it work of separation}
\wsep\ of the interface. It does not refer to chemical equilibrium
states and therefore does not involve chemical
potentials of the separate components:

\begin{eqnarray} 
W_{sep}&=&(G_{SAO}+G_{SB}-G_{int})/2S \nonumber\\
       &=&\gamma_{AO}+\gamma_B-\gamma_{int}. \label{wsep}
\end{eqnarray} 

For brevity we do not explicitly indicate the temperature and pressure
dependence of all the terms unless it needs to be emphasised.  An
important point to note about this quantity, which makes it relatively
straightforward to calculate, is that the separate slabs of metal and
oxide have exactly the same composition as the two slabs which are
joined to form an interface. This would not in general be the case if
these slabs and the interface were in equilibrium with a given
environment (constant $\mu_i$), because one would expect for example
some loss or gain of surface oxygen or metal from the oxide to the
vapour phase when the surfaces are created. If the interface as well as
the exposed surfaces are the ones which are in chemical equilibrium
(which we denote by superfix $eq$), an equation similar to (\ref{wsep})
defines the {\it work of adhesion}:

\begin{equation} 
W_{ad}= \gamma_{AO}^{eq}+\gamma_B^{eq}-\gamma_{int}^{eq}. \label{wad}
\end{equation} 

This is the quantity of relevance to contact angles and wetting for
example, and unlike \wsep\ it is not obtainable by a simple
comparison of three total energies.

Calculations of \wsep\ for Nb/\al2o3 were reported in our
Letter\cite{Finnis1} and these have been extended
here, as described in the following sections. \wsep\ is probably
more relevant than $W_{ad}$ in formulating a fracture criterion, when
internal surfaces are formed which are not in equilibrium, but in order
to predict the equilibrium structure of interfaces we also need to be able to
evaluate Eqns.(\ref{gammaint})-(\ref{gammaB}).

We now introduce the quantity $g_{AO}$, the Gibbs energy per formula
unit of bulk \AO~ in equilibrium with metal $A$ and oxygen in vapour form:
\begin{equation} 
g_{AO}(T,P)=m\mu_A(T,P)+n\mu_O(T,P), \label{gAO}
\end{equation}  
so that
\begin{equation} 
G_{AO} = (N_A/m)g_{AO} \label{scaleg}
\end{equation} 
is the Gibbs energy of a stoichiometric cell containing $N_A$ atoms of
$A$. Inserting (\ref{gAO}) and (\ref{excess}) into (\ref{gammaAO}) gives
the surface energy of the oxide in a form which makes the effect
of the excess oxygen explicit:  
\begin{equation} 
\gamma_{AO}=(G_{SAO}-G_{AO})/2S
           -\Gamma_O\mu_O. \label{gammaAO2}
\end{equation} 

Consider now how to go about calculating the two surface energies from
(\ref{gammaB}) and (\ref{gammaAO2}), which we will eventually combine
with \wsep\ in (\ref{wsep}) to give us $\gamma_{int}$. The Gibbs
energy of all slabs can be calculated at $T=0K$ and $P=0$ from first principles, it
is just the total energy.  If the slabs are bulk pure material, their
Gibbs energy can be corrected to temperature {T} by using experimental
specific heat data.  On the other hand when the slabs are separated in
the supercell by a layer of vacuum to represent free surfaces, there is
no such experimental data and the correction to finite $T$ could
 be done by calculating the phonon spectrum and using the
quasiharmonic approximation for the free energy. This has been done
previously for classical ionic models by Taylor and
coworkers\cite{Taylor}, in order to obtain the temperature dependence
of their surface energy, but we have not yet made the equivalent
calculation with our {\it ab initio} code. For a metal slab (Ag), the
quasiharmonic free energy based on {\it ab initio} phonon frequencies
was recently calculated by Xie and coworkers\cite{Xie}. In the case of
the pure metal slab, the chemical potential $\mu_B$ is the Gibbs energy
per atom of a bulk slab. The surface energy of $B$ is therefore
obtained from the results of two supercell total energy calculations in
the standard way.  The main present issue, which is less familiar in
the context of total energy calculations, is how to calculate the
significant term due to the chemical potential of oxygen, which must be
included when the surface of the oxide is non-stoichiometric ($\Gamma_O
\ne 0$).  

The chemical potential of oxygen is well described in terms of its partial 
pressure $P_{O_2}$ by the standard ideal gas expression 
\begin{equation} 
\mu_O=\mu_O^0+\frac{1}{2}kT\,\log (P_{O_2}/P^0). \label{muO}
\end{equation} 
In Eqn.~(\ref{muO}), $\mu_O^0$ is
the oxygen chemical potential in its standard state (STP) at 
$T^0$=298.15K, $P^0$=1at. Chemists would set $\mu_O^0$ to zero by
definition, but we cannot do that since our zero of energy is already
defined as the energy of separated ions and electrons at $T=0$K. On the
other hand the energy of oxygen molecules is not something we want to
calculate, since there are well known problems in using density
functional theory for this system. Fortunately, we can circumvent 
the problem by using a
thermodynamic cycle. From the defining equation for 
the standard Gibbs energy of formation $\Delta G^0_{AO}$: 
\begin{equation} 
g_{AO}^0=m\mu_{A}^0+n\mu_O^0+\Delta G_{AO}^0,\label{deltag}
\end{equation}
we obtain the troublesome oxygen chemical potential at STP in terms of
$g_{AO}^0$ and $\Delta G_{AO}^0$. The quantities $g_{AO}^0$ and
$\mu_{A}^0$ are things we {\it can} calculate accurately, and we can
look up $\Delta G^0_{AO}$ in tables of thermodynamic data.

Inserting $\mu_O^0$ from (\ref{deltag}) into (\ref{muO}) and (\ref{muO}) 
into (\ref{gammaAO2})
gives us our final expression for the surface energy of the oxide:
\begin{eqnarray} 
\gamma_{AO}&=&(G_{SAO}(T,P)-\frac{N_A}{m}g_{AO}(T,P))/2S \nonumber\\ 
           &&-\Gamma_O(g_{AO}^0 -m\mu_{A}^0-\Delta G^0_{AO})/n\nonumber\\ 
           &&-\Gamma_O\frac{1}{2}kT\,\log (P_{O_2}/P^0). \label{gammaAO3}
\end{eqnarray} 
from which we obtain the final expression for the interfacial energy by subsituting
(\ref{gammaAO3}) into (\ref{wsep}): 
\begin{eqnarray} 
\gamma_{int}&=&\gamma_B(T,P)-W_{sep}(T,P)\nonumber\\ 
 &&+(G_{SAO}(T,P)-\frac{N_A}{m}g_{AO}(T,P))/2S\nonumber\\ 
 &&-\Gamma_O(g_{AO}^0-m\mu_{A}^0-\Delta G^0_{AO})/n\nonumber\\ 
 &&-\Gamma_O\frac{1}{2}kT\,\log (P_{O_2}/P^0). \label{gammaint2}
\end{eqnarray} 
The quantities $g_{AO}^0$ and $\mu_A^0$ entering the third line of
(\ref{gammaint2}) are well described  by $T=0$K quantities which we
calculate. It can be verified that correcting them to standard state
has a negligible effect on the surface energy.

The minimum physically meaningful value of $P_{O_2}$, which we denote
$P_{O_2}^{min}$, is set by the condition that if $P_{O_2} \leq
P_{O_2}^{min}$ the oxide would spontaneously decompose into metal and
oxygen.  Neglecting the small variation in solid energies with
temperature by comparison with $\Delta G^0_{AO}$ this condition is:
\begin{equation} \log (P_{O_2}^{min}/P^0)=\frac{2}{nkT}\Delta G^0_{AO}.
\label{po2min} \end{equation} 
Similarly, the maximum physically
meaningful value of $P_{O_2}$ is defined by the lowest standard Gibbs
energy of formation of a metal B oxide $\Delta G^O_{BO}$:
\begin{equation} \log (P_{O_2}^{max}/P^0)=\frac{2}{n^\prime kT}\Delta
G^0_{BO} \label{po2max} \end{equation} where the first oxide to form
would have the stoichiometry $B_{m^\prime}O_{n^\prime}$. The
thermodynamic data used here are summarised in Table~\ref{Table1}.

\section{Method of total energy calculation}

\begin{figure}[h] 
\begin{center} 
\leavevmode 
\psfig{file=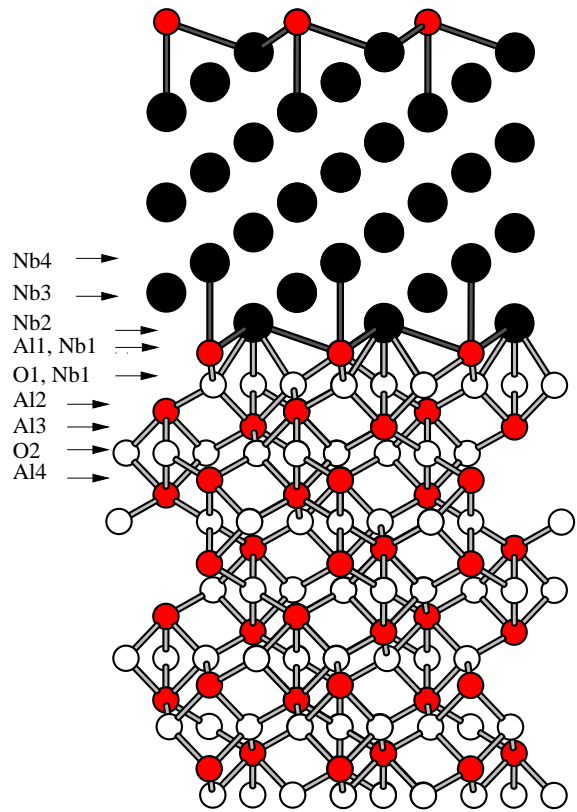,width=7.9 in} 
\end{center}
\caption{Side view of the
Nb(111)/Al$_2$O$_3$(0001) interface, showing labelling of the layers. }\label{Fig1} 
\end{figure}

For the interface calculations we use the total energy plane wave
pseudopotential method based on Lanczos diagonalization of the
Kohn-Sham density matrix\cite{Alavi}.  The supercell has the form of a
rhombohedral prism and in the stoichiometric slab it contains 45
atoms:  14 Al, 21 O and 10 Nb atoms (see Fig.\ref{Fig1}). By stripping
off the outer plane of Al from each interface we obtain an interface
which is O-terminated with an O excess $\Gamma_O \cdot S=+1.5$ atoms
per surface unit cell.  By adding the surface plane of Al atoms to the
neutral interface we make an oxygen poor interface, with the negative O
excess $\Gamma_O \cdot S=-1.5$. The total energy of the contents of a
supercell is minimized with respect to the atomic coordinates by the
quasi-Newton method with Hessian updated using the
Broyden-Fletcher-Goldfarb-Shano (BFGS) method. The pseudopotential for Nb was
of Troullier-Martins form  \cite{Troullier}, with $s$ and $d$
non-locality. The pseudopotential for O was of optimised form \cite{Lee}, with $s$
non-locality. The pseudopotential for Al was of Gonze type \cite{Gonze}
with $s$ non-locality. 

 All calculations were made with two
$k$-points in the irreducible wedge of the Brillouin zone, and with a
plane-wave cut-off of 40 Ry. The effect of increasing the plane wave
cutoff from 40 to 60Ry was to reduce \wsep\ by 3.3\% for the Nb/Al
interface, which we take as a satisfactory indication of the basis set
convergence. For the neutral 45 atom interface we have made test
calculations with six $k$-points which results in a decrease of total
energy by about 4mRy and very small ($<10^{-3}$\,nm ) changes of relaxed
positions of atoms compared with two $k$-point calculations.  The
effect of increasing the $k$-point sampling from 2 points to 9 is to
change \wsep\ by less than 1\%.

By doubling the original unit cell in the x-y plane we obtained a 180
atom cell, with which we recalculated the wavefunctions at the gamma point with
the previously relaxed atomic coordinates. The gamma point wavefunctions in this cell were
used for Mulliken population analysis which was
made by projecting the optimized wave functions onto the pseudoatomic
orbitals $|\phi_{i\alpha}>$ (i labels site, $\alpha$ - orbitals)
according to the procedure suggested in\cite{Sanchez95}. The
``spillage'' of each occupied orbital $\psi$ was less than 1.5\%.

\section{Relaxation of the interface}

\begin{figure}[h] 
\begin{center} 
\leavevmode 
\psfig{file=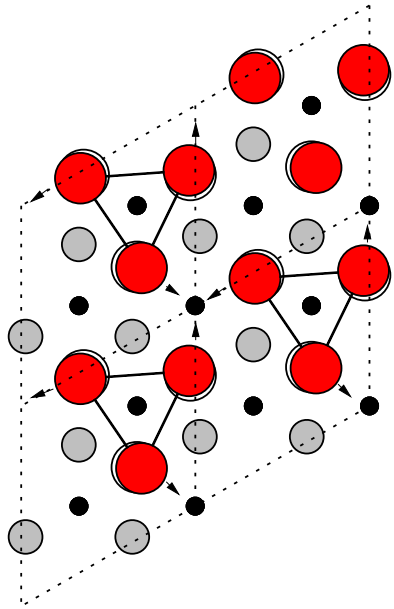,width=2.9 in} 
\end{center}
\caption{Plane view of the (0001) surface of neutral
alumina showing the lateral relaxation within the topmost O plane. The
rotation and expansion of the O triangle below the surface Al atom is
indicated by arrows. }\label{Fig2}
\end{figure}

The slab with which the Nb(111)/\al2o3(0001) interface was modelled is
shown in Fig.\ref{Fig1}.  The interlayer relaxation of the interface
has been reported previously\cite{Finnis1}, and we refer to that paper
for results. Here we mention a feature which has not previously been
discussed, namely the relaxations parallel to the interface, which we
refer to as in-plane relaxations. It has been found that to make a
calculation of the interlayer relaxation of the alumina surface one
needs to take into account the in-plane relaxations of the oxygen
atoms\cite{Finnis1}, which were neglected in some earlier
work\cite{Kruse2}.  The present results show that in-plane relaxation
of the oxygen ions is a general feature of the structure near the
interface.  The geometry of these relaxations is described by two
parameters; the rotation $\alpha$ and the bond length extension $\delta
r$ of the equilateral triangle of oxygen atoms in a plane, illustrated
in Fig.\ref{Fig2}. The rotation and dilation of this equilateral
triangle does not break any symmetry, preserving for example the three
fold axis about the centre of the triangle concerned.

The calculated values of $\alpha$ and $\delta r$ are shown in
Table~\ref{Table2} for the terminating and second layer oxygen planes
(denote by subscripts 1 and 2) in five cases. The first two cases,
labelled N(b)/A(O) and N(b)/A(Al) are the  O and Al terminated  bulk
Nb/\al2o3 interfaces. Cases three and four are pure alumina surfaces,
labelled A(O) and A(Al) to indicate that they are oxygen and aluminium
terminated respectively. Case five, labelled N(m)/A(O) is a Nb
monolayer on an oxygen terminated alumina surface.  One can see that
the in-plane relaxation is a feature of all the systems studied.  From
the evidence of the first two layers, the rotation of O-triangles and
the increase of the O-O bond lengths appears to be localised near the
surface of the alumina. The Nb monolayer on the O-terminated surface of
alumina shows this effect most strongly, which is quite surprising,
since the interplanar relaxation in this case is much less than that of
the Al layer for which the Nb substitutes. There is experimental
confirmation of the effect, obtained by small angle X-ray
diffraction\cite{Guenard}, in the case of the A(Al) surface.  In this
case the experimental results are $\delta r_1=4.5\%$ and
$\alpha_1$=$6.7^{ \circ}$, compared with our calculated results of
$\delta r_1=3.2\%$ and
$\alpha_1$=$3.1^{\circ}$. The agreement is only qualitative.

\section{Work of separation} 

All our results for the calculated work of separation \wsep\ of
different interfaces and cleavage planes are shown in
Table~\ref{Table3}.  The column of `unrelaxed' results refers to values
obtained by assuming bulk unrelaxed atomic positions both at the
interface and for the free surfaces. The interplanar spacing between O
and Nb across the interface in this case was simply taken as the bulk
spacing between O and Al planes. The results in the `relaxed' column
are calculated with atomic positions relaxed both before and after
cleavage.

The effect of relaxations on \wsep\ naturally depends on the
interface.  It is most pronounced when an Al terminated \al2o3 surface
is exposed, because of the large relaxation of this surface, which
lowers its surface energy by about 1.5\,Jm$^{-2}$. On the other hand, if
the relaxation of the {\it interface} dominates the energy balance in
Eqn.(\ref{wsep}) then the relaxed value of \wsep can even be larger
than the unrelaxed value, as in the case of cleavage between Nb and O
at the N(b)/A(O) interface.

The lowest value of \wsep , 2.7\,Jm$^{-2}$, is found for the cleavage of
bulk Nb from the stoichiometric Al-terminated alumina. The highest
values are found for the cleavage of bulk Nb from the O-terminated
alumina surface. Indeed we can deduce from Table~\ref{Table3} that
this interface would be unlikely to separate between Nb and O planes, but would
prefer to separate inside the Nb, leaving a monolayer of Nb on the surface, or
even between O and Al, leaving a monolayer of O on the Nb surface.

The highest value (relaxed) of all in Table I is for
the cleavage of pure \al2o3 between O and Al planes. Experimentally,
$\alpha-$\al2o3 does not cleave on the basal plane at all, but its
lowest energy cleavage on this plane would clearly be between Al
planes.  This is what one expects on the basis of charge neutrality
arguments, because by cleaving between Al planes two identical, neutral
surfaces are created.  On the other hand by cleaving between O and Al,
different surfaces are created which, in order to be neutral, require
the oxygen or aluminium at the surface to be in an unfavourable valence
state, hence this is a final state of especially high energy. 

The above interpretations are supported by the Mulliken populations
shown in Table~\ref{Table4} for the three interfaces between bulk Nb
and \al2o3.  We make the usual caveat here that Mulliken charges do not
have absolute significance, since they depend on the choice of basis
set, but they are nevertheless a useful indicator of trends in ionicity
or covalency. The interfaces in Table III are labelled by their oxygen
excesses, to highlight certain trends with the stoichiometry of the
interface. Bulk oxygen carries a Mulliken charge of -1, and for the
oxygen plane nearest the interface this value is reduced to -0.99, -0.93 and
-0.86 in turn as the excess of oxygen at the interface is increased
from negative to positive. The change is rather modest, indicating that
oxygen does not readily alter its valence state. The charge on the
interfacial oxygen is provided by the terminating layer of Al in the
case of Al termination, or in the case of the oxygen terminated
interface the electrons are provided mainly by the first two layers of
Nb.

\section{Interfacial free energy and oxygen pressure}

Five surface energies are shown in Fig.\ref{Fig3}a as a function of
$P_{O_2}$. The x-axis is appropriate to a temperature of 1500K; to obtain the
results at temperature $T$ the numbers on the x-axis should be multiplied by
$1500/T$. The O-rich ($\Gamma_O \cdot S=1.5$) and O-poor ($\Gamma_O \cdot S=-1.5$)
alumina surfaces have negative and positive slopes respectively, while
the stoichiometric \al2o3 and pure Nb(111) surface energies are constant,
and by chance nearly equal. The most negatively sloping surface energy
we have plotted here refers to the Nb(111) surface with an attached monolayer of
oxygen. It becomes negative at an oxygen pressure inside the regime of
stability of NbO. 

The interfacial free energy from Eqn.(\ref{gammaint}) is shown as a
function of $P_{O_2}$ in Fig.\ref{Fig3}c for three interfaces, O-rich,
stoichiometric and O-poor ($\Gamma_O \cdot S=1.5,0,-1.5$). To generate
the interfacial free energies one has to subtract the work of
separation shown for convenience in Fig.\ref{Fig3}b, from the sum of
the equilibrium surface energies of the two corresponding free
surfaces. With increasing $P_{O_2}$ the O-rich interface becomes
increasingly stable, the Al-rich interface less stable and the free
energy of the stoichiometric interface remains constant, exactly
parallel to the behaviour of the free surfaces. The interfaces can only
be in thermodynamic equilibrium in the range of oxygen pressure which
is indicated on the figure. At 1500K this range is as given in Table
\ref{Table1}; at values of $P_{O_2}$ above this range, the Nb would
oxidise to NbO, and below it the alumina would decompose.

The {\it work of adhesion} at a given $P_{O_2}$ can also be estimated
from the results on this graph using Eqn.(\ref{wad}).  The free surface
and interface energies should be those with lowest free energy at the
given oxygen pressure, and these can be read off from Figs.\ref{Fig3}a
and \ref{Fig3}c.  The result is plotted in Fig.\ref{Fig3}d.

\section{Discussion and Conclusions} 

We have made a careful distinction between work of separation, a
mechanically defined quantity, and work of adhesion, a thermodynamic
quantity, focusing on how to go about calculating these quantities
within an atomistic model. We particularly consider the interface
between a metal and an oxide, since it is of practical importance and
since oxygen is a troublesome component for which to calculate the
chemical potential, a key quantity in interfacial energies.  A useful
practical equation for the free energy of an interface involving oxygen
has been derived, namely Eqn.(\ref{wad}), which gets around the
previous difficulty by using a thermodynamic cycle to express the
result in terms of quantities which can be readily calculated, namely
the total energies of slabs, and quantities which can be obtained from
tables, namely the standard Gibbs energy of formation of the oxide and
the Gibbs energies of the bulk materials relative to their $T=0$K
values.

To illustrate and apply the method we have made a number of first
principles calculations for  Nb(111)/\al2o3 (0001) interfaces,
oxygen rich, oxygen poor and stoichiometric, and for
several surfaces. We fully relax the atomic positions in supercells
using a plane wave, pseudopotential methodology. The relaxations are
significant, and in all cases they involve in-plane as well as
interlayer relaxations of the oxygen ions. Results on the work of
separation of these interfaces were given in a Letter\cite{Finnis1}
recently, and we have extended them to include the possibility of a
cleavage of the O-terminated interface which leaves the Nb coated with
oxygen. This turns out indeed to be a lower energy mode of separation
(4.9\,Jm$^{-2}$) than the alternative which leaves an oxygen rich
\al2o3 surface behind (9.8\,Jm$^{-2}$), because the favourable degree
of ionicity of oxygen is thereby preserved as it is in both  bulk
alumina and its stoichiometric surface. Considering further the strongly bound
O-terminated Nb/\al2o3 interface, it turns out that the hypothetical
processes of (i) cleavage within bulk Nb (4.2\,Jm$^{-2}$), or (ii)
leaving a monolayer of Nb on the oxide surface (3.8\,Jm$^{-2}$),  or
even (iii) cleavage within bulk \al2o3 (3.9\,Jm$^{-2}$) are all
marginally of lower energy than the cleavage which takes oxygen with
the niobium. 

By combining the results of our calculations with thermodynamic data
we obtain surface energies and interfacial energies as a function of
oxygen partial pressure and temperature. An approximation we make
here is in omitting the temperature dependence of the solid state free
energy, but we include the $kT \log (P_{O_2}/P^0)$ term which
describes the temperature dependence of the oxygen chemical potential;
this is also the term which describes the dependence of all the
interfacial and surface free energies on oxygen pressure. It is clear
how a more accurate calculation could be made by implementing the
quasiharmonic approximation to correct solid surface free energies, and
it will probably become a routine matter to include such a correction
in future work. Another approximation is made by considering only a small
set of possible interface and surface compositions which we think are
representative. Nevertheless, despite the present simplifications, some
clear results have emerged.

Of the free surfaces of \al2o3 , the stoichiometric one, terminated by
a single layer of Al, is the most stable over the whole range of oxygen
partial pressure up to over one atmosphere. It may be that a treatment
of the temperature dependence of the energy of the slabs could modify
the upper and lower bounds on pressure somewhat. Correction of the LDA
error is also likely to lower surface energies by 10-20\% (I. G.
Batyrev, unpublished).  For example, work of C. E.  Sims {\it et
al}~\cite{Sims} with classical potentials indicates that the surface
energies of Al-terminated \al2o3 can be reduced by up to  0.2-0.3
Jm$^{-2}$ at 1500-2000K.  We expect an oxygen terminated surface to be
stable at a pressure not too far from atmospheric, but we cannot
unfortunately be more quantitative in the prediction at this stage.  At
very low oxygen pressures it is also reasonable that the experimentally
observed Al-rich ($\sqrt{31} \times \sqrt{31}$) structure is stable; we
cannot model a supercell of the size needed to calculate this. Instead
we modelled a much simpler Al-rich interface, which is predicted to
become the most stable one just above $P_{O_2}^{min}$ where \al2o3
decomposes. Since the experimental $\sqrt{31} \times \sqrt{31}$ is a
very Al-rich surface ($\Gamma_O \cdot S = -7.5$ in the present
notation), the slope of its surface energy versus $\log (P_{O_2}/P^0)$
is correspondingly very steep and positive, and it must intersect all the
other surface energies just above $P_{O_2}^{min}$.

\begin{figure} 
\begin{center} 
\leavevmode 
\psfig{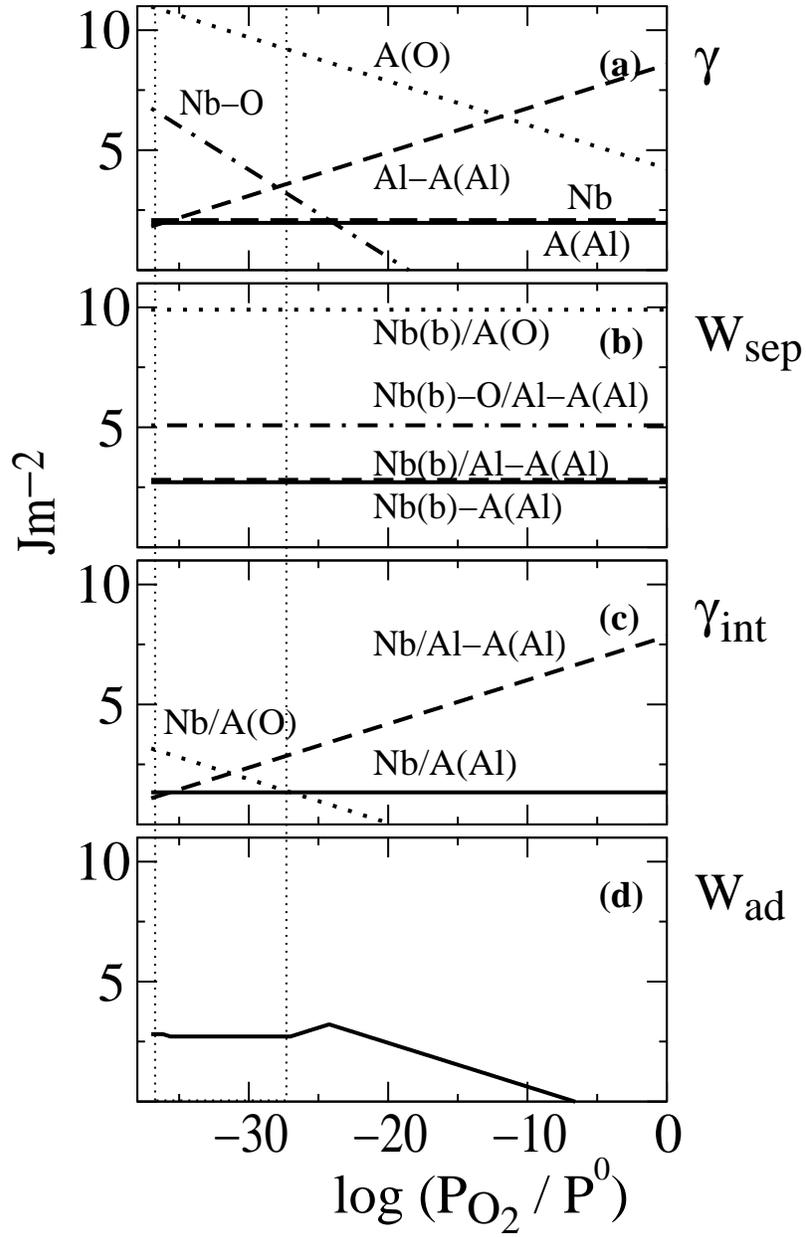} 
\end{center}
\caption{(a) Surface energies as a 
function of oxygen partial pressure at 1500 K. (b) Works of separation $W_{sep}$ (see Table
III). (c) Interfacial energies.  (d) Work of adhesion $W_{ad}$, obtained by 
subtracting equilibrium interfacial energy from equilibrium surface energies.
The region to the left between the vertical
lines corresponds to the possible equilibrium states of the interface.}\label{Fig3}
\end{figure}

The Nb free surface should obviously become unstable with respect to
some adsorption of oxygen when $P_{O_2} > P_{O_2}^{max}$, the pressure
at which NbO begins to form. The particular configuration and
concentration of an oxygen monolayer which we have calculated is not
likely to be the optimum configuration of the first oxygen covered
Nb(111) surface, but it does become more stable than the free surface
at pressures somewhat above $P_{O_2}^{max}$ (Fig.\ref{Fig3}a).

A significant new result is the theoretical analysis of the {\it
thermodynamic} stability of the O-terminated interface, the strong
bonding of which we discussed above. No interface is thermodynamically
stable above the (very low) oxygen pressure at which NbO forms, but
over most of the range below this the O-terminated interface is less
stable than the Al-terminated one (Fig.\ref{Fig3}c), despite its strong
bonding. In fact at the very lowest pressure of oxygen, as would pertain in the
presence of pure aluminium, our prediction is of an Al-enriched
interface. The experimental indications from EELS\cite{Bruley} show no
evidence for Al-Nb bonding, and suggest rather the existence of the
O-terminated interface. According to our analysis this could only be
marginally in thermodynamic equilibrium if the oxygen pressure is being
`buffered' by Nb/NbO and lies close to $P_{O_2}^{max}$, which does not
seem unreasonable. 

We have not included in our comparison interfaces with
a different macroscopic orientation such as the
Nb(110)/\al2o3 (0001) interface~\cite{Mayer1}. Although this interface
is believed to be thermodynamically more stable than the Nb(111)/\al2o3 (0001)
interface, the kinetic barrier to changing the macroscopic orientation
is presumably much greater than the barriers to changing the local
interface structure. 

Although the formalism has been developed for describing metal-oxide
bonding, there are obvious applications to systems in which water or
other substances may contaminate surfaces or interfaces. The comparison
of the energetics of interfaces with differing amounts of segregation
follows the same lines. The application of the present formalism
using a thermodynamic cycle to avoid the most difficult calculations
may be fruitful in other situations in the field of interface chemistry.

\acknowledgements{We thank J. Hutter for technical help with the
calculations.  This work has been supported by the UK Engineering and
Physical Sciences Research Council under grants No. GR/L08380 and
GR/M01753, and by the European Communities HCM Network ``Electronic
Structure Calculations of Materials Properties and Processes for
Industry and Basic Science'' under grant No.  ERBFMRXCT980178. The
Centre for Supercomputing in Ireland is gratefully acknowledged for
computer resources.}

\pagebreak

\begin{table} [h]
\caption{Thermodynamic data used for calculating the dependence of
surface and interfacial energies on oxygen partial pressure. The pressure
represents the dissociation pressure of the oxide at 1500K. $P^0$ is one
standard atmosphere. }
\begin{center}
\begin{tabular}{||c|c|c||}
  & $\alpha-$\al2o3  &   NbO    \\ \hline
\ $\Delta G^0$ (kJ/mol) \cite{CRC}& -391.9 & -378.6 \\
\ $\log_{10} (P_{O_2}/P^0)$ &   -36.8    &   -26.4   \\
\end{tabular}
\end{center}
\label{Table1}
\end{table}

\begin{table}[h]

\caption{In-plane relaxation of O and Al terminated interfaces and
surfaces of alumina. $\alpha$ and $\delta r$ correspond to the angle of
rotation and bond length increase of O
triangles; surface and subsurface layers are indicated by the subscripts. }
\begin{center} 
\begin{tabular}{||l|r|r|r|r|r||}
	  & N(b)/A/(0)&N(b)/A(Al)&A(O)&A(Al)&N(m)/A(O)  \\ \hline
\ $\alpha_1$ &4.1      & 4.2         &4.1   &3.1&8.4      \\ 
\ $\alpha_2$ &0.1      & 0.2         &0.5   &0.2& 0.4     \\ 
\ $\delta r_1$     &4.2      & 4.3         &4.3   &3.2&8.4      \\
\ $\delta r_2$ &0.2 & 0.2         &0.2   &0.5&0.6
 
\end{tabular} 
\end{center} 
\label{Table2}
\end{table}

\begin{table} 
\caption{\wsep (in $J/m^{2}$) for both unrelaxed and
relaxed structures. } 
\begin{center} 
\begin{tabular}{||l|c|cc||} \ 
Interface      &
Cleavage plane                                 &Unrel & Rel \\ \hline
\ N(b)/N(b)      & ...Nb-Nb-Nb $\wr$ Nb-Nb-Nb...                & 4.9
& 4.2  \\ 
\ A(Al)/A(Al)    & ...Al-O-Al $\wr$
Al-O-Al...                  & 7.0 & 3.9  \\ 
\ A(Al)/A(O)     &
...Al-Al-O $\wr$ Al-Al-O...                  &13.3 &12.7 \\
\ N(m)/A(O)      & \hskip 30pt Nb $\wr$ O-Al-Al...              & 10.9
&10.8 \\ 
\ N(b)/A(O)      & ...Nb-Nb-Nb $\wr$
O-Al-Al-O...               & 9.3 &9.8 \\ 
\ N(b)/A(Al)     & ...Nb-Nb-Nb
$\wr$ Al-O-Al-Al...              & 4.2 & 2.7 \\ 
\ N(b)/Nb-A(Al) &
...Nb-Nb-Nb $\wr$ Nb-O-Al-Al ...             & 4.0  &3.8\\
\ N(b)/Al-A(Al)  & ...Nb-Nb-Nb $\wr$ Al-Al-O...                 & 4.4
& 2.8\\ 
\ N(b)-O/Al-A(Al)& ...Nb-Nb-O $\wr$ Al-Al-O...
& 7.3  & 4.9 
\end{tabular} 
\end{center}
\label{Table3} 
\end{table}

\begin{table} 
\caption{Mulliken charges for atoms near the interface as
a function of the excess of O. The notation for atomic planes is as in
Fig.~1. } 
\begin{center} 
\begin{tabular}{||l|rrr||} 
\ $\Gamma_O \cdot S$
&1.5  & 0 &-1.5\\ \hline 
\ Nb2  &0.27       &0.37   &-0.13   \\
\ Nb1/Al1&0.77     &0.73   &0.36  \\ 
\ O1   &-0.86    &-0.93   &missing   \\
\ Al2  &1.49       &1.50   &0.15      \\ 
\ Al3  &1.52      &1.51 &0.95      \\ 
\ O2   &-1.00      &-1.00   &-0.99 
\end{tabular}
\end{center} 
\label{Table4} 
\end{table} 

\onecolumn

\end{document}